\documentclass[%
 reprint,
 amsmath,amssymb,
prb,
]{revtex4-2}

\usepackage{latexsym}
\usepackage{graphicx}
\usepackage{siunitx}
\usepackage{xcolor}
\usepackage{soul}
\usepackage[capitalise]{cleveref}
\begin{document}
\title{Local Control of Supercurrent Density in Epitaxial Planar Josephson Junctions}

\author{Bassel Heiba Elfeky$^{1}$} 
\altaffiliation{These authors contributed equally.}

\author{Neda Lotfizadeh$^{1}$} 
\altaffiliation{These authors contributed equally.}

\author{William F. Schiela$^{1}$}

\author{William M. Strickland$^{1}$}

\author{Matthieu Dartiailh$^{1}$} 

\author{Kasra Sardashti$^{1}$}

\author{Mehdi Hatefipour$^{1}$}

\author{Peng Yu$^{1}$}

\author{Natalia Pankratova$^{2}$}

\author{Hanho Lee$^{2}$}

\author{Vladimir E. Manucharyan$^{2}$}

\author{Javad Shabani$^{1}$}
\email{jshabani@nyu.edu}

\affiliation{%
$^{1}$Department of Physics, New York University, New York, NY 10003 USA\\
$^{2}$Department of Physics, Joint Quantum Institute, and Quantum Materials Center, University of Maryland, College Park, MD 20742, USA}

\date{\today}

\begin{abstract}

\textbf{The critical current response to an applied out-of-plane magnetic field in a Josephson junction provides insight into the uniformity of its current distribution. In Josephson junctions with semiconducting weak links, the carrier density and, therefore the overall current distribution could be modified electrostatically via metallic gates. Here, we show local control of the current distribution in an epitaxial Al-InAs Josephson junction equipped with five mini-gates. We demonstrate that not only can the junction width be electrostatically defined but also we can locally adjust the current profile to form superconducting quantum interference devices. Our studies show enhanced edge conduction in such long junctions, which can be eliminated by mini-gates to create a uniform current distribution.}

\end{abstract}


\maketitle



Recent advancements in epitaxial growth of superconductors on semiconductors aiming to create an abrupt and uniform interface have enabled the fabrication of Josephson junctions with unprecedented favorable properties \cite{shabani_two-dimensional_2016,mayer_superconducting_2019, bottcher_superconducting_2018, mayer_gate_2020, krogstrup_epitaxy_2015, kjaergaard_transparent_2017, lee_transport_2019}. Recently, it has been shown that these Josephson junctions (JJs) can host Majorana fermions \cite{hell_two-dimensional_2017, pientka_topological_2017} by the application of a Zeeman field \cite{dartiailh_phase_2021} or a phase bias \cite{ren_topological_2019,fornieri_evidence_2019}. Aside from the stabilization of the topological state, the system needs to be amenable to reconfiguration, as suggested by Alicea et al. \cite{alicea_non-abelian_2011}, to realize such behavior. This reconfiguration is difficult to realize using the conventional single top gate covering the entire junction, which modifies the total carrier density while only inexplicitly affecting the current distribution. Even though the local control of the supercurrent distribution is an essential element for the functionality of these systems, it has not been experimentally demonstrated thus far.

In this work, we study JJs fabricated on epitaxial Al-InAs.  Such devices have shown high transparency \cite{mayer_superconducting_2019, kjaergaard_transparent_2017, dartiailh_missing_2021},  a spin-orbit induced anomalous phase \cite{mayer_gate_2020}, a superconducting-insulating transition \cite{bottcher_superconducting_2018} and topological superconductivity \cite{ dartiailh_phase_2021, fornieri_evidence_2019}. 
We introduce a design with an array of top gates arranged evenly along the junction, referred to as mini-gates (MGs), to control and study the local supercurrent density. Using the MG design, by suppressing the carrier density in specific parts of the junction while keeping the other parts active, we tune the supercurrent to effectively change the width of the JJ or create patterns of superconducting quantum interference devices (SQUIDs). By applying an out-of-plane magnetic field, a Fraunhofer pattern is observed in the critical current and is used to obtain the spatial supercurrent density distribution through a well-known process of supercurrent density reconstruction established by Fulton et al. \cite{dynes_supercurrent_1971}(See Supplementary Materials for details). The setup is conveniently configurable and could be utilized to create topological and non-topological areas in a JJ by changing the spin-orbit coupling locally \cite{mayer_gate_2020, dartiailh_phase_2021, wickramasinghe_transport_2018}, in a similar fashion to earlier proposals with nanowires \cite{alicea_non-abelian_2011, bauer_dynamics_2018} and recently proposed fusion experiments in JJs with local gate control \cite{zhou_fusion_2021}.

\begin{figure*}[ht!]
    \centerline{\includegraphics[width=1.0\textwidth]{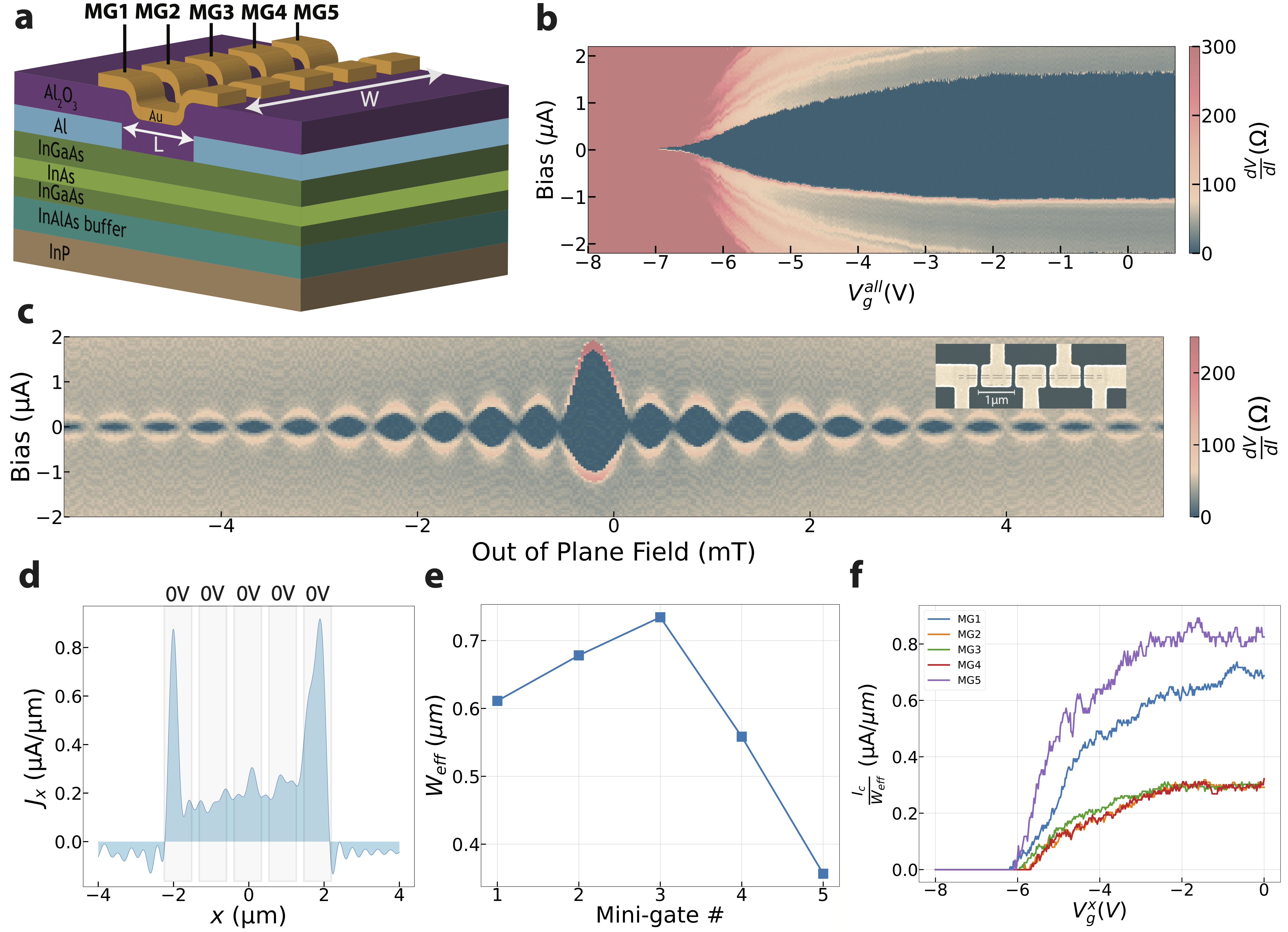}}
    \caption{\textbf{a} A schematic diagram of the JJ heterostructure with five top Au mini-gates (MGs), labeled MG1 through MG5, on a junction of length $L$ and width $W$. The superconducting contacts are made of Al and the QW consists of a layer of InAs grown between two layers of In$_{0.81}$Ga$_{0.19}$As. \textbf{b} Differential resistance as a function of the bias current and applied gate voltage where all the gates are driven together, $V^{all}_{g} = V^{1,2,3,4,5}_{g}$. \textbf{c} Differential resistance as a function of the bias current and out-of-plane magnetic field with all the MGs at $V^{all}_{g} = 0$ V. Inset: A false-color SEM image of the five MGs. The grey lines outline the junction (not visible in the SEM image) under the MGs. \textbf{d} Calculated supercurrent density distribution across the junction obtained from the interference pattern in \textbf{c}. Each MG's expected spatial coverage is defined by shaded sections and the value on top of each section indicates the applied gate voltage $V^{i}_{g}$ to each MG. \textbf{e} Effective width W$_{eff}$ governed by each MG calculated from five different interference maps where only one MG is set to $V^{i}_{g} = 0$V while the rest are set to $V^{x \neq i} = -8$V (see Supp. Materials for all interference  maps). \textbf{f} Critical current per unit effective width $\frac{I_{c}}{W_{eff}}$ controlled by each individual MG. Here, only one MG is swept from $V^{i}_{g} = +0.7$V to $-8$V at a time, while others are kept at $V^{x \neq i} = -8$V.}
    \label{fig:fig_intro}
\end{figure*}

Our heterostructure is grown by molecular beam epitaxy on a semi-insulating InP (100) substrate followed by a graded buffer layer. To create an optimal interface and a high mobility 2DEG, a 7 nm layer of InAs is grown between two layers of In$_{0.81}$Ga$_{0.19}$As. The 2DEG is then capped with a layer of in-situ grown thin epitaxial Al, and the junction is formed using a selective wet etch of Al. A layer of Al$_{2}$O$_{3}$ is deposited subsequently using atomic layer deposition followed by an array of five Ti/Au top mini-gates (MGs). The schematic drawing of the heterostructure and the junction is shown in \cref{fig:fig_intro}a where the MGs are numbered MG1 through MG5. The dimensions of the JJ device studied is $\SI{4}{\micro m}$ wide $W$ and $\SI{150}{\nano m}$ long $L$, with the MGs being separated by approximately $\SI{125}{\nano m}$. The inset in \cref{fig:fig_intro}c shows a false-color scanning electron microscopy (SEM) image of the five MGs. The junction area covered by each MG is approximately the same and each MG can be tuned separately to change the local carrier density of a segment of the junction by applying a gate voltage (V$_{g}$) that alters the critical current of that segment. All the measurements in this work are performed at T $\sim 30$ mK. Magnetotransport measurements were performed on Hall bars formed on the InAs 2DEG for characterization. Details of these measurements can be found in the Supplemental Materials as well as more details on the fabrication process and the measurement setup.

By applying a negative gate voltage to the MGs, the critical current under each section covered by the MGs can be suppressed and ultimately drive the device to a normal resistive state (by activating all MGs). \cref{fig:fig_intro}b shows the differential resistance as a function of the gate voltage $V^{all}_{g}$ and current bias. Here, the same gate voltage is applied to all the MGs simultaneously, $V^{all}_{g} = V^{1,2,3,4,5}_{g}$, where the superscript indicates the specific MG. It can be seen that at $V^{all}_{g} = -8$V, there is no supercurrent flowing, and the device overall is in the resistive regime. 

Applying an out of plane magnetic field $B_{\bot}$ while keeping all the MGs at zero voltage, $V^{all}_{g} = 0$V, the critical current exhibits a familiar Fraunhofer-like form seen in the differential resistance map plotted in \cref{fig:fig_intro}c. The maximum critical current is seen to be $I^{max}_{c} = \SI{1.8}{\micro A}$ and the normal resistance $R_{n} = \SI{323}{\Omega}$ is obtained by fitting the linear high bias part of the IV curve (shown in supplementary figure S1). The large ratio between $\mathrm{e} I_{c}R_{n}$ and $\Delta_{Al}$ indicates a high transparency interface, where  $\Delta_{Al} = \SI{216}{\micro eV}$ is the superconducting gap of Al as calculated from the measured critical temperature. The periodicity of the interference pattern and $B_{\bot} A=\Phi_0$ can be used to calculate the effective area of the junction to be $A_{eff} = \SI{5.31}{\mu}m^{2}$ which is larger than the nominal geometry due to flux focusing caused by the Meissner effect and London penetration \cite{suominen_anomalous_2017,ghatak_anomalous_2018}. 

The interference pattern in \cref{fig:fig_intro}c shows lifted nodes with the critical current not reaching precisely zero for the first couple of nodes. The origin of such node lifting was theoretically predicted to be due to the presence of a supercurrent contribution from Majorana bound states\cite{potter_anomalous_2013} and attributed to non-sinusoidal contributions to the current-phase relation \cite{kurter_evidence_2015, williams_unconventional_2012}. Without an in-plane magnetic field applied, our junction is expected to be topologically trivial, and given the absence of an even-odd effect, it is most likely that the node-lifting in our device is due to an inhomogeneous current distribution. Other trivial effects such as the presence of disorder across the junction and screening effects have been discussed in literature \cite{hui_proximity-induced_2014}; however, our device has a $l_{e} = \SI{268}{nm}$ compared to length of the junction, $L = \SI{150}{\nano m}$, should be near the short ballistic regime consistent with earlier studies \cite{mayer_superconducting_2019}

The spatial distribution of the current density can be reconstructed (see Supplementary Materials for method details) from the interference pattern shown in \cref{fig:fig_intro}c. \cref{fig:fig_intro}d plots the current density distribution across the junction, where $x$ is the spatial position along the JJ's width centered around $x = 0$. The distribution is shown to be, in fact, inhomogeneous, with the contribution of the supercurrent being more significant towards the edges appearing in the form of two edge channels. To understand the origin of these edge channels, characterizing each segment of the junction using the MG is necessary. 

\begin{figure*}[ht!]
    \centerline{\includegraphics[width=1.0\textwidth]{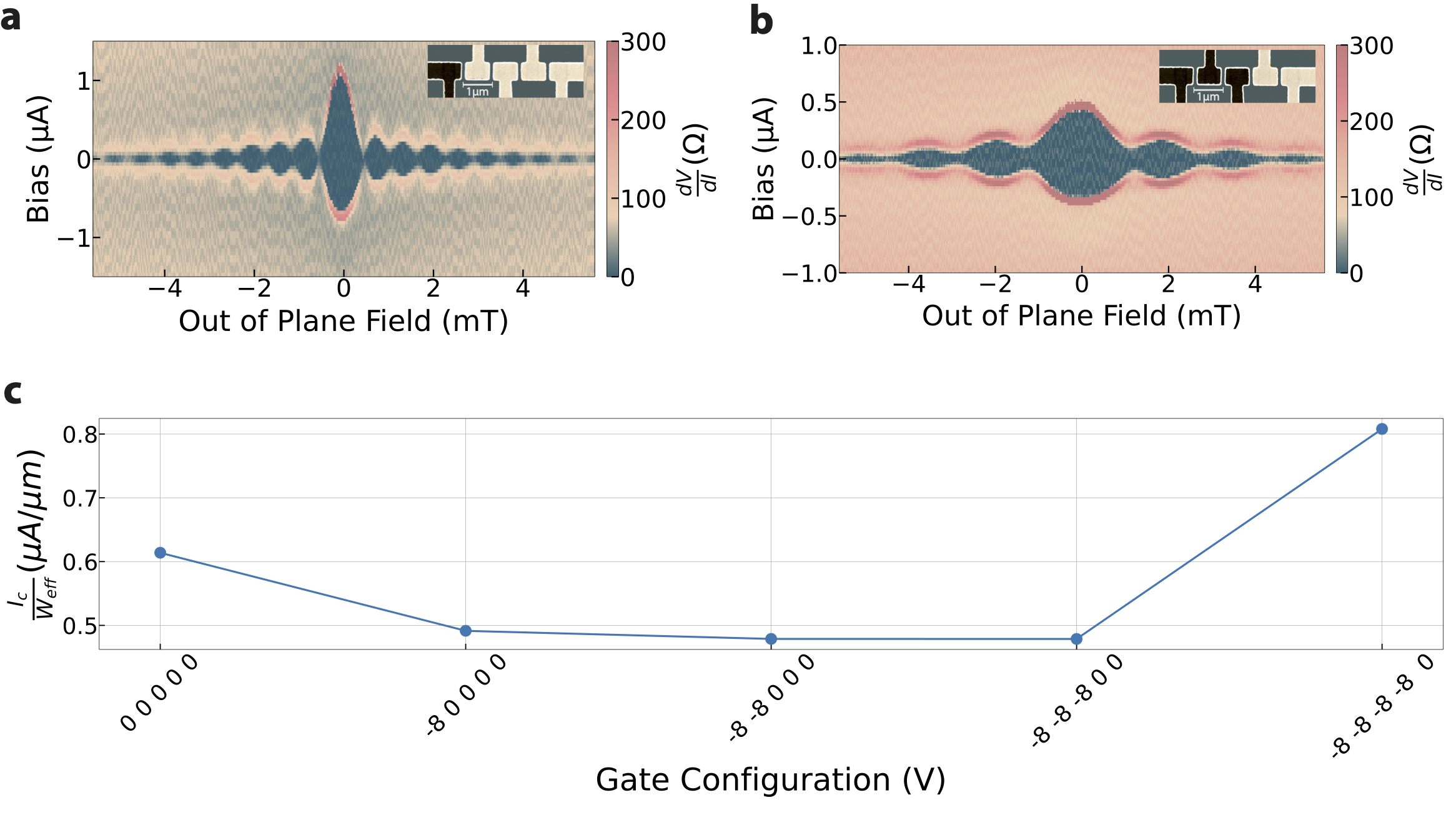}}
    \caption{Interference maps for two different gate configurations that represent reducing the effective width of the junction to \textbf{a} $\frac{4}{5}$ and \textbf{b} $\frac{2}{5}$ the original width by sequentially suppressing the supercurrent from one side of the junction. A complete set of the interference maps for different effective widths is shown in Supp Fig.S2. The inset shows a false-color SEM image of the five mini-gates (MG1 being the leftmost one) and indicates which parts of the junction are driven to a resistive state by applying $V^{i}_{g} = -8$V (shaded black) and which are superconducting $V^{i}_{g} = 0$V (shaded white).  \textbf{c} Critical current per unit effective width $\frac{I_{c}}{W_{eff}}$ for the specified gate configurations, where the numbers correspond to the gate voltage values applied in volts to MG1-MG5 (left to right), that represent reducing the effective width of the junction. The width $W_{eff}$ is the cumulative effective width covered by each MG that is at $V^{i}_{g} = 0$V from \cref{fig:fig_intro}e.}
    \label{fig:fig_jj_ends}
\end{figure*}

The mini-gate design can be utilized to study different sections of the junction separately by keeping only one MG at $V^{i}_{g} = 0$V while driving the other four MGs at $V^{x \neq i}_{g} = -8$V to drive the rest of the junction to the normal resistive mode, e.g.\ $V^{1}_{g} = 0$V and $V^{2,3,4,5}_{g} = -8$ V. The corresponding periodicity of the interference patterns (complete set of interference maps shown in Supp. Fig S2) can be used to find the effective width W$_{eff}$ governed by each MG as plotted in \cref{fig:fig_intro}e. We can further sweep $V^{i}_{g}$ applied to the MG controlling the superconducting section while keeping the rest of the junction in a resistive state and find the amount of $I_c$ controlled by each MG. \cref{fig:fig_intro}f indicates that the three inner MGs (MG2, MG3, and MG4) control approximately the same amount of critical current per unit effective width $\frac{I_{c}}{W_{eff}}$. In comparison, the outer two MGs have a larger share of the current density. This agrees with the current density distribution in \cref{fig:fig_intro}d extracted from the interference pattern observed with $V^{all}_{g} =0$.

The characteristics of a JJ, including its current distribution, are subject to change when the effective width of the junction is altered. This can be achieved using the MG design by suppressing the supercurrent sequentially from the left section of the junction. First, we apply $V^{1}_g = -8$V to MG1 and keep the other four MGs at zero gate voltage. We find the width of the junction to now be approximately $\SI{3.0}{\micro m}$, which is close to the nominal value of $\SI{3.1}{\micro m}$. Then, by applying $V^{1,2}_g = -8$V while keeping the rest at zero gate voltage, the width of the junction decreases to approximately $\SI{2.25}{\micro m}$. This process can be repeated with MG3 and MG4 to decrease the effective width of the junction further. Two of the interference patterns for these gate configurations representing the reduction of the effective width of the junction to $\frac{4}{5}$ and $\frac{2}{5}$ of the original width are shown in \cref{fig:fig_jj_ends}a, b and the complete set in Supp. Fig. S3.

The critical current per unit effective width $\frac{I_{c}}{W_{eff}}$, where $W_{eff}$ is the total effective width of the superconducting sections, for gate configurations corresponding to different effective JJ widths is shown in \cref{fig:fig_jj_ends}c. $\frac{I_{c}}{W_{eff}}$ changes as a function of the number of resistive sections, with the first and last gate configuration exhibiting the larger $\frac{I_{c}}{W}$ as expected due to enhanced edge conductance. The three middle gate configurations do not show a significant change in $\frac{I_{c}}{W}$. We note that changing the effective width of the junction is desirable if the system is driven into a topological regime. Recently, sub-gap state energy oscillations were observed \cite{albrecht_exponential_2016} due to the overlap of hybridized Majorana wavefunctions on multiple nanowires and interpreted as enhanced protection of zero modes as a function of increasing wire length. In our geometry, this could be done in-situ.

\begin{figure}[ht!]
    \centerline{\includegraphics[width=0.5\textwidth]{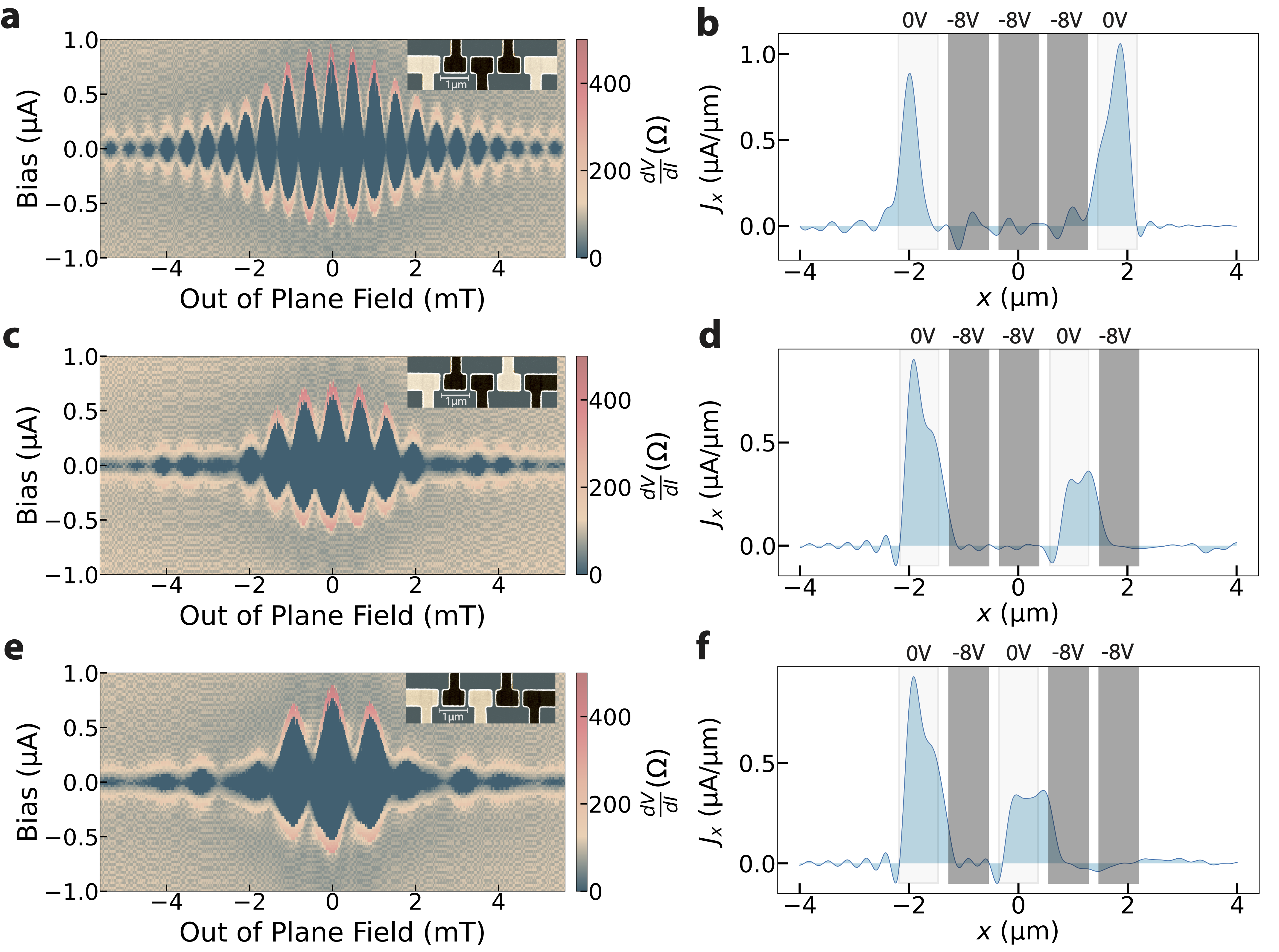}}
    \caption{\textbf{a,c,e} SQUID-like oscillations for three different gate configurations where only two sections of the junction are superconducting. The inset of each figure shows a false-color SEM image of the five MGs where the shaded black ones are driven to $V^{i}_{g} = -8$V, and the shaded white ones are kept at $V^{i}_{g} = 0$V. \textbf{b,d,f} The corresponding reconstructed current profiles of each gate configuration. Shaded sections define each MG's expected coverage and the value on top of each section indicates the value of the applied gate voltage $V^{i}_{g}$ to each MG.
    }  
    \label{fig:fig_squid}
\end{figure}

The process of local suppression of supercurrent can be further extended to obtain SQUID-like behavior from a usual straight junction by creating a resistive area between two superconducting ones using the MGs. \cref{fig:fig_squid} shows the changes in the interference pattern and their corresponding current profile for different SQUID MG configurations where the interspace between the superconducting regions decreases. 
There is a notable difference in the oscillations of \cref{fig:fig_squid}a, c and e representing the superconducting sections being separated by three, two, and one MG widths, respectively. The configuration with the largest separation between the superconducting regions and most symmetric current density distribution in \cref{fig:fig_squid}a looks more like a conventional SQUID. \cref{fig:fig_squid}c and e, which have similar asymmetric current distributions, show similar dependence on the separation of the superconducting regions. Further, the node-lifting in \cref{fig:fig_squid}c, and e is more pronounced due to a more asymmetric current distribution compared to the one observed in \cref{fig:fig_intro}d, which implies that the node-lifting is due to an inhomogeneous current distribution rather than any non-trivial effects.



\begin{figure*}[ht!]
    \centerline{\includegraphics[width=1.0\textwidth]{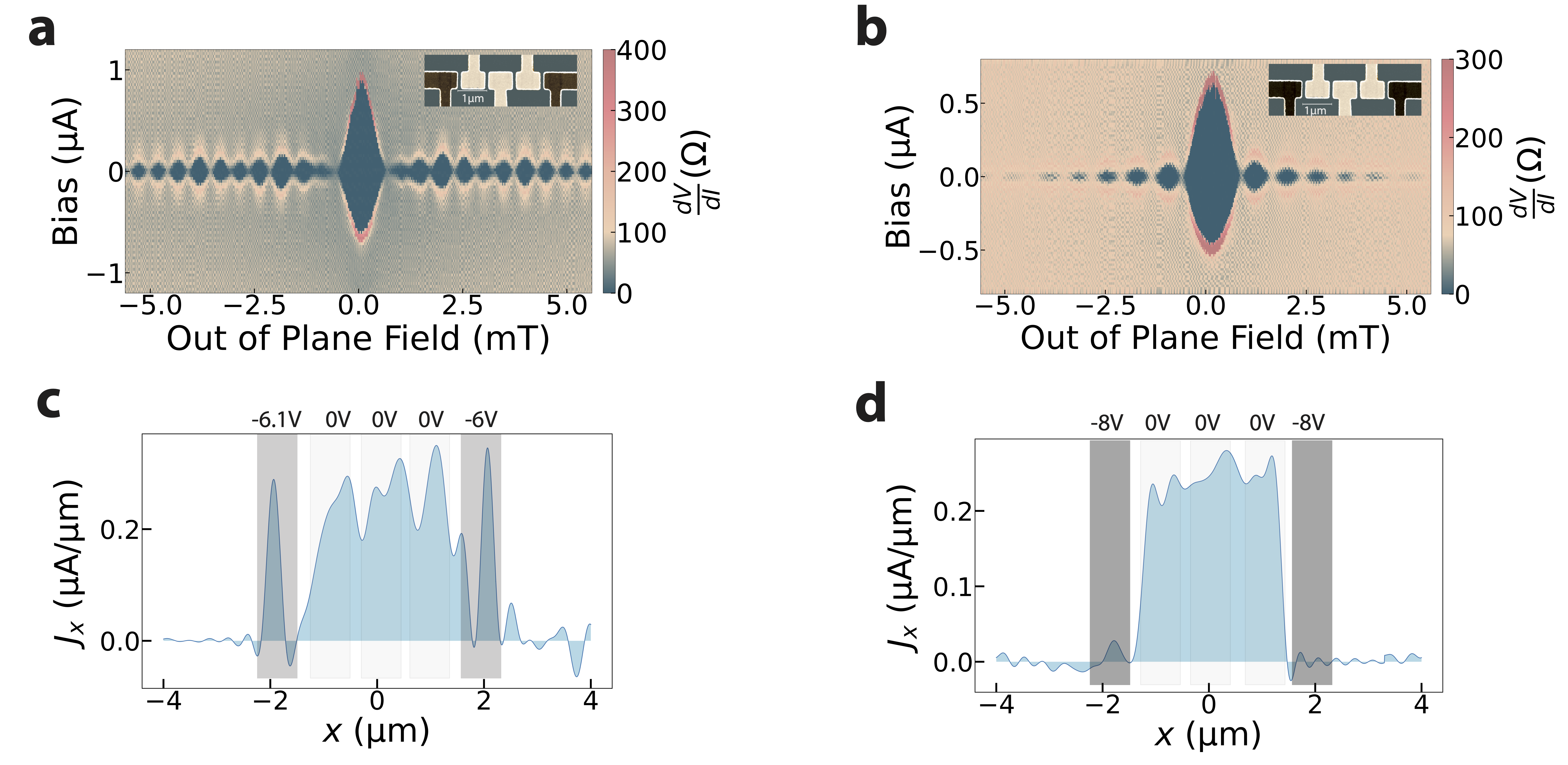}}
    \caption{\textbf{a,c} Interference pattern and reconstructed current profile corresponding to leveling outer peaks with central current distribution by applying $V^{1}_{g} = -6.1$V and $V^{5}_{g} = -6$V while keeping the three middle MGs at  $V^{1,2,3}_{g} = 0$V. \textbf{b,d} Interference pattern and reconstructed current profile corresponding to suppressing the supercurrent in both outer areas of the junction to achieve a homogeneous current distribution. The inset of each interference measurement figure shows a false-color SEM image of the five MGs, shaded to indicate $V_{g}^{i}$. Shaded sections in the current distribution plots define each MG's expected coverage and the value on top of each section indicates the value of the applied gate voltage $V^{i}_{g}$ to each MG.
    } 
    \label{fig:fig_jj_edge}

\end{figure*}

In general, the supercurrent density distribution along the junction is an important factor that can provide insight into a junction's characteristics. In two-dimensional systems, such as topological insulators, topological superconductivity is expected to arise in the form of edge states \cite{hasan_colloquium_2010}. Still, topologically trivial edge/surface states have shown to arise in specific systems \cite{qi_topological_2011,ghatak_anomalous_2018,de_vries_h_2018,deVries_insb_flake_2019}. Superconducting edge transport has also been studied in JJs formed on other structures such as quantum spin Hall insulators \cite{hart_induced_2014}, nanoplates \cite{ying_magnitude_2020},type-II semimetal \cite{huang_edge_2020}, graphene \cite{zhu_edge_2017}, and 2DEGs of InAs \cite{de_vries_h_2018}, all in the presence of a single top and/or bottom gates. The supercurrent density profiles in these cases have shown large edge peaks at the ends of the junctions similar to those observed in our device in \cref{fig:fig_intro}d. However, the origin of the edge accumulated current density in the previous studies was attributed mostly to topological edge channels, strong spin-orbit coupling, non-trivial topology of the gapped Dirac spectrum, or crossed Andreev reflection. Since none of these properties are present in our system, the origin of the edge states in our junction should be different from these previous cases.

In systems such as ours, the Schottky barrier height and the Fermi level position depend on the semiconductor-metal interface and details of fabrication. Specifically, the Fermi level, pinned at the semiconductor surface, is usually related to the crystal direction and surface treatments \cite{canali_low-temperature_1998}. For most of the group III-V and IV semiconductors, the Fermi level pinning is independent of the metal and is below the center of the band gap \cite{mead_fermi_1964}. However, in the case of InAs, the Fermi level could be pinned above the conduction band minimum \cite{bhargava_fermi-level_1997}. This results in band-bending effects \cite{suominen_anomalous_2017,ying_magnitude_2020,hui_proximity-induced_2014} and electron accumulation towards the ends of the junction. We performed 2D Poisson-Schr\"{o}dinger calculations on a $\SI{4}{\micro m}$ junction, similar to the one being studied and simulated the electron density $\rho$ and conduction band energy $E$ as a function of position (Fig. S7 in Supplementary Material). This calculation assumes a fixed pinning position \cite{bhargava_fermi-level_1997} of $-130$ meV at the surface and enforces it at the InAs-air interface. The band bending effects result in two narrow edge peaks, each about about $\SI{30}{\nano m}$ wide, in the electron density consistent with enhanced conductance at the ends of the junction observed in \cref{fig:fig_intro}d. Details on the calculations can be found in Supplementary Materials. It can be seen in \cref{fig:fig_intro}d that MG1 and MG5 encompass the prominent peaks in current density and can be tuned to neutralize such edge effects. We note that experimentally we have not observed this current accumulation in smaller width junctions, with examples shown in Fig. S4 in the Supplementary Materials. 

To neutralize the enhanced edge current density, one can achieve a uniform distribution by lowering the supercurrent on the edges using MG1 and MG5.  Increasing the applied voltages to $V^{1}_{g} = -6.1$V and $V^{5}_{g} = -6$V to equalize the edge peak heights with the central homogeneous distribution can still however lead to a discontinuous current distribution, as plotted in \cref{fig:fig_jj_edge}c.
This can be explained by the fact that the outer MGs encompass more than just the width of the peaks. When $V^{1}_{g}$ and $V^{5}_{g}$ are significant enough to level out the peaks, the rest of the space covered by those two MGs has reached a low critical current or a resistive state. From the current reconstruction the FWHM is seen to be about $400-\SI{600}{\nano m}$ which is larger than the FWHM estimated from the band bending calculations suggesting a trivial edge of about $\SI{30}{\nano m}$. This discrepancy is expected due to limited sampling in the current reconstruction process as well as the effects of finite temperature and field dispersion on the measurements.

To completely remove the edge channels and establish a homogeneous current distribution, the two ends of the junction under the outermost MGs could be driven to the normal resistive mode by applying $V^{1,5}_{g} = -8$V. By keeping the three middle MGs at 0V, the supercurrent can evenly flow through the inner region of the junction. In this case, the device is expected to behave as a narrower JJ consistent with the data in \cref{fig:fig_jj_edge}b, which is similar to a conventional Fraunhofer pattern. The node-lifting is not observed in this interference pattern indicating a homogeneous current distribution as seen in the reconstructed current profile shown in \cref{fig:fig_jj_edge}d.



In conclusion, by applying an out-of-plane magnetic field to an epitaxial Al-InAs JJ with five top MGs, we have extracted the spatial current distribution, which exhibits enhanced edge current density for our $\SI{4}{\micro m}$ wide junction. Using the five MGs distributed along the JJ, we demonstrate how the local current distribution can be utilized to change the effective width of the JJ and in the creation of SQUID-like configurations, evident from the corresponding interference pattern and constructed current distribution. Simple 2D Poisson-Schr\"{o}dinger calculations suggest that the enhanced edge current density can be attributed to band-bending effects resulting in electron accumulation towards the edges of the junction. Our mini-gate design can be utilized to suppress these edge effects, resulting in a homogeneous current distribution across the junction. The ability to locally tune the current distribution width of the junction could open a path for studying topological superconductivity in Josephson junctions.



{\bf Acknowledgments -} NYU team acknowledges support by DARPA TEE award no. DP18AP900007 and UMD team acknowledges DARPA young faculty award no. D17AP00025.  W.~F.~S. acknowledges funding from the NDSEG Fellowship.

\bibliography{Ref}

\providecommand{\noopsort}[1]{}\providecommand{\singleletter}[1]{#1}%
\begin{thebibliography}{36}%
\makeatletter
\providecommand \@ifxundefined [1]{%
 \@ifx{#1\undefined}
}%
\providecommand \@ifnum [1]{%
 \ifnum #1\expandafter \@firstoftwo
 \else \expandafter \@secondoftwo
 \fi
}%
\providecommand \@ifx [1]{%
 \ifx #1\expandafter \@firstoftwo
 \else \expandafter \@secondoftwo
 \fi
}%
\providecommand \natexlab [1]{#1}%
\providecommand \enquote  [1]{``#1''}%
\providecommand \bibnamefont  [1]{#1}%
\providecommand \bibfnamefont [1]{#1}%
\providecommand \citenamefont [1]{#1}%
\providecommand \href@noop [0]{\@secondoftwo}%
\providecommand \href [0]{\begingroup \@sanitize@url \@href}%
\providecommand \@href[1]{\@@startlink{#1}\@@href}%
\providecommand \@@href[1]{\endgroup#1\@@endlink}%
\providecommand \@sanitize@url [0]{\catcode `\\12\catcode `\$12\catcode
  `\&12\catcode `\#12\catcode `\^12\catcode `\_12\catcode `\%12\relax}%
\providecommand \@@startlink[1]{}%
\providecommand \@@endlink[0]{}%
\providecommand \url  [0]{\begingroup\@sanitize@url \@url }%
\providecommand \@url [1]{\endgroup\@href {#1}{\urlprefix }}%
\providecommand \urlprefix  [0]{URL }%
\providecommand \Eprint [0]{\href }%
\providecommand \doibase [0]{https://doi.org/}%
\providecommand \selectlanguage [0]{\@gobble}%
\providecommand \bibinfo  [0]{\@secondoftwo}%
\providecommand \bibfield  [0]{\@secondoftwo}%
\providecommand \translation [1]{[#1]}%
\providecommand \BibitemOpen [0]{}%
\providecommand \bibitemStop [0]{}%
\providecommand \bibitemNoStop [0]{.\EOS\space}%
\providecommand \EOS [0]{\spacefactor3000\relax}%
\providecommand \BibitemShut  [1]{\csname bibitem#1\endcsname}%
\let\auto@bib@innerbib\@empty
\bibitem [{\citenamefont {Shabani}\ \emph {et~al.}(2016)\citenamefont
  {Shabani}, \citenamefont {Kjaergaard}, \citenamefont {Suominen},
  \citenamefont {Kim}, \citenamefont {Nichele}, \citenamefont {Pakrouski},
  \citenamefont {Stankevic}, \citenamefont {Lutchyn}, \citenamefont
  {Krogstrup}, \citenamefont {Feidenhans'l}, \citenamefont {Kraemer},
  \citenamefont {Nayak}, \citenamefont {Troyer}, \citenamefont {Marcus},\ and\
  \citenamefont {Palmstrøm}}]{shabani_two-dimensional_2016}%
  \BibitemOpen
  \bibfield  {author} {\bibinfo {author} {\bibfnamefont {J.}~\bibnamefont
  {Shabani}}, \bibinfo {author} {\bibfnamefont {M.}~\bibnamefont {Kjaergaard}},
  \bibinfo {author} {\bibfnamefont {H.~J.}\ \bibnamefont {Suominen}}, \bibinfo
  {author} {\bibfnamefont {Y.}~\bibnamefont {Kim}}, \bibinfo {author}
  {\bibfnamefont {F.}~\bibnamefont {Nichele}}, \bibinfo {author} {\bibfnamefont
  {K.}~\bibnamefont {Pakrouski}}, \bibinfo {author} {\bibfnamefont
  {T.}~\bibnamefont {Stankevic}}, \bibinfo {author} {\bibfnamefont {R.~M.}\
  \bibnamefont {Lutchyn}}, \bibinfo {author} {\bibfnamefont {P.}~\bibnamefont
  {Krogstrup}}, \bibinfo {author} {\bibfnamefont {R.}~\bibnamefont
  {Feidenhans'l}}, \bibinfo {author} {\bibfnamefont {S.}~\bibnamefont
  {Kraemer}}, \bibinfo {author} {\bibfnamefont {C.}~\bibnamefont {Nayak}},
  \bibinfo {author} {\bibfnamefont {M.}~\bibnamefont {Troyer}}, \bibinfo
  {author} {\bibfnamefont {C.~M.}\ \bibnamefont {Marcus}},\ and\ \bibinfo
  {author} {\bibfnamefont {C.~J.}\ \bibnamefont {Palmstrøm}},\ }\bibfield
  {title} {\bibinfo {title} {Two-dimensional epitaxial
  superconductor-semiconductor heterostructures: {A} platform for topological
  superconducting networks},\ }\href
  {https://doi.org/10.1103/PhysRevB.93.155402} {\bibfield  {journal} {\bibinfo
  {journal} {Physical Review B}\ }\textbf {\bibinfo {volume} {93}},\ \bibinfo
  {pages} {155402} (\bibinfo {year} {2016})}\BibitemShut {NoStop}%
\bibitem [{\citenamefont {Mayer}\ \emph {et~al.}(2019)\citenamefont {Mayer},
  \citenamefont {Yuan}, \citenamefont {Wickramasinghe}, \citenamefont {Nguyen},
  \citenamefont {Dartiailh},\ and\ \citenamefont
  {Shabani}}]{mayer_superconducting_2019}%
  \BibitemOpen
  \bibfield  {author} {\bibinfo {author} {\bibfnamefont {W.}~\bibnamefont
  {Mayer}}, \bibinfo {author} {\bibfnamefont {J.}~\bibnamefont {Yuan}},
  \bibinfo {author} {\bibfnamefont {K.~S.}\ \bibnamefont {Wickramasinghe}},
  \bibinfo {author} {\bibfnamefont {T.}~\bibnamefont {Nguyen}}, \bibinfo
  {author} {\bibfnamefont {M.~C.}\ \bibnamefont {Dartiailh}},\ and\ \bibinfo
  {author} {\bibfnamefont {J.}~\bibnamefont {Shabani}},\ }\bibfield  {title}
  {\bibinfo {title} {Superconducting proximity effect in epitaxial al-{InAs}
  heterostructures},\ }\href {https://doi.org/10.1063/1.5067363} {\bibfield
  {journal} {\bibinfo  {journal} {Applied Physics Letters}\ }\textbf {\bibinfo
  {volume} {114}},\ \bibinfo {pages} {103104} (\bibinfo {year}
  {2019})}\BibitemShut {NoStop}%
\bibitem [{\citenamefont {Bøttcher}\ \emph {et~al.}(2018)\citenamefont
  {Bøttcher}, \citenamefont {Nichele}, \citenamefont {Kjaergaard},
  \citenamefont {Suominen}, \citenamefont {Shabani}, \citenamefont
  {Palmstrøm},\ and\ \citenamefont {Marcus}}]{bottcher_superconducting_2018}%
  \BibitemOpen
  \bibfield  {author} {\bibinfo {author} {\bibfnamefont {C.~G.~L.}\
  \bibnamefont {Bøttcher}}, \bibinfo {author} {\bibfnamefont {F.}~\bibnamefont
  {Nichele}}, \bibinfo {author} {\bibfnamefont {M.}~\bibnamefont {Kjaergaard}},
  \bibinfo {author} {\bibfnamefont {H.~J.}\ \bibnamefont {Suominen}}, \bibinfo
  {author} {\bibfnamefont {J.}~\bibnamefont {Shabani}}, \bibinfo {author}
  {\bibfnamefont {C.~J.}\ \bibnamefont {Palmstrøm}},\ and\ \bibinfo {author}
  {\bibfnamefont {C.~M.}\ \bibnamefont {Marcus}},\ }\bibfield  {title}
  {\bibinfo {title} {Superconducting, insulating and anomalous metallic regimes
  in a gated two-dimensional semiconductor–superconductor array},\ }\href
  {https://doi.org/10.1038/s41567-018-0259-9} {\bibfield  {journal} {\bibinfo
  {journal} {Nature Physics}\ }\textbf {\bibinfo {volume} {14}},\ \bibinfo
  {pages} {1138} (\bibinfo {year} {2018})}\BibitemShut {NoStop}%
\bibitem [{\citenamefont {Mayer}\ \emph {et~al.}(2020)\citenamefont {Mayer},
  \citenamefont {Dartiailh}, \citenamefont {Yuan}, \citenamefont
  {Wickramasinghe}, \citenamefont {Rossi},\ and\ \citenamefont
  {Shabani}}]{mayer_gate_2020}%
  \BibitemOpen
  \bibfield  {author} {\bibinfo {author} {\bibfnamefont {W.}~\bibnamefont
  {Mayer}}, \bibinfo {author} {\bibfnamefont {M.~C.}\ \bibnamefont
  {Dartiailh}}, \bibinfo {author} {\bibfnamefont {J.}~\bibnamefont {Yuan}},
  \bibinfo {author} {\bibfnamefont {K.~S.}\ \bibnamefont {Wickramasinghe}},
  \bibinfo {author} {\bibfnamefont {E.}~\bibnamefont {Rossi}},\ and\ \bibinfo
  {author} {\bibfnamefont {J.}~\bibnamefont {Shabani}},\ }\bibfield  {title}
  {\bibinfo {title} {Gate controlled anomalous phase shift in al/{InAs}
  josephson junctions},\ }\href {https://doi.org/10.1038/s41467-019-14094-1}
  {\bibfield  {journal} {\bibinfo  {journal} {Nature Communications}\ }\textbf
  {\bibinfo {volume} {11}},\ \bibinfo {pages} {212} (\bibinfo {year}
  {2020})}\BibitemShut {NoStop}%
\bibitem [{\citenamefont {Krogstrup}\ \emph {et~al.}(2015)\citenamefont
  {Krogstrup}, \citenamefont {Ziino}, \citenamefont {Chang}, \citenamefont
  {Albrecht}, \citenamefont {Madsen}, \citenamefont {Johnson}, \citenamefont
  {Nygård}, \citenamefont {Marcus},\ and\ \citenamefont
  {Jespersen}}]{krogstrup_epitaxy_2015}%
  \BibitemOpen
  \bibfield  {author} {\bibinfo {author} {\bibfnamefont {P.}~\bibnamefont
  {Krogstrup}}, \bibinfo {author} {\bibfnamefont {N.~L.~B.}\ \bibnamefont
  {Ziino}}, \bibinfo {author} {\bibfnamefont {W.}~\bibnamefont {Chang}},
  \bibinfo {author} {\bibfnamefont {S.~M.}\ \bibnamefont {Albrecht}}, \bibinfo
  {author} {\bibfnamefont {M.~H.}\ \bibnamefont {Madsen}}, \bibinfo {author}
  {\bibfnamefont {E.}~\bibnamefont {Johnson}}, \bibinfo {author} {\bibfnamefont
  {J.}~\bibnamefont {Nygård}}, \bibinfo {author} {\bibfnamefont
  {C.}~\bibnamefont {Marcus}},\ and\ \bibinfo {author} {\bibfnamefont {T.~S.}\
  \bibnamefont {Jespersen}},\ }\bibfield  {title} {\bibinfo {title} {Epitaxy of
  semiconductor–superconductor nanowires},\ }\href
  {https://doi.org/10.1038/nmat4176} {\bibfield  {journal} {\bibinfo  {journal}
  {Nature Materials}\ }\textbf {\bibinfo {volume} {14}},\ \bibinfo {pages}
  {400} (\bibinfo {year} {2015})}\BibitemShut {NoStop}%
\bibitem [{\citenamefont {Kjaergaard}\ \emph {et~al.}(2017)\citenamefont
  {Kjaergaard}, \citenamefont {Suominen}, \citenamefont {Nowak}, \citenamefont
  {Akhmerov}, \citenamefont {Shabani}, \citenamefont {Palmstrøm},
  \citenamefont {Nichele},\ and\ \citenamefont
  {Marcus}}]{kjaergaard_transparent_2017}%
  \BibitemOpen
  \bibfield  {author} {\bibinfo {author} {\bibfnamefont {M.}~\bibnamefont
  {Kjaergaard}}, \bibinfo {author} {\bibfnamefont {H.}~\bibnamefont
  {Suominen}}, \bibinfo {author} {\bibfnamefont {M.}~\bibnamefont {Nowak}},
  \bibinfo {author} {\bibfnamefont {A.}~\bibnamefont {Akhmerov}}, \bibinfo
  {author} {\bibfnamefont {J.}~\bibnamefont {Shabani}}, \bibinfo {author}
  {\bibfnamefont {C.}~\bibnamefont {Palmstrøm}}, \bibinfo {author}
  {\bibfnamefont {F.}~\bibnamefont {Nichele}},\ and\ \bibinfo {author}
  {\bibfnamefont {C.}~\bibnamefont {Marcus}},\ }\bibfield  {title} {\bibinfo
  {title} {Transparent {Semiconductor}-{Superconductor} {Interface} and
  {Induced} {Gap} in an {Epitaxial} {Heterostructure} {Josephson} {Junction}},\
  }\href {https://doi.org/10.1103/PhysRevApplied.7.034029} {\bibfield
  {journal} {\bibinfo  {journal} {Physical Review Applied}\ }\textbf {\bibinfo
  {volume} {7}},\ \bibinfo {pages} {034029} (\bibinfo {year}
  {2017})}\BibitemShut {NoStop}%
\bibitem [{\citenamefont {Lee}\ \emph {et~al.}(2019)\citenamefont {Lee},
  \citenamefont {Shojaei}, \citenamefont {Pendharkar}, \citenamefont
  {{McFadden}}, \citenamefont {Kim}, \citenamefont {Suominen}, \citenamefont
  {Kjaergaard}, \citenamefont {Nichele}, \citenamefont {Zhang}, \citenamefont
  {Marcus},\ and\ \citenamefont {Palmstrøm}}]{lee_transport_2019}%
  \BibitemOpen
  \bibfield  {author} {\bibinfo {author} {\bibfnamefont {J.~S.}\ \bibnamefont
  {Lee}}, \bibinfo {author} {\bibfnamefont {B.}~\bibnamefont {Shojaei}},
  \bibinfo {author} {\bibfnamefont {M.}~\bibnamefont {Pendharkar}}, \bibinfo
  {author} {\bibfnamefont {A.~P.}\ \bibnamefont {{McFadden}}}, \bibinfo
  {author} {\bibfnamefont {Y.}~\bibnamefont {Kim}}, \bibinfo {author}
  {\bibfnamefont {H.~J.}\ \bibnamefont {Suominen}}, \bibinfo {author}
  {\bibfnamefont {M.}~\bibnamefont {Kjaergaard}}, \bibinfo {author}
  {\bibfnamefont {F.}~\bibnamefont {Nichele}}, \bibinfo {author} {\bibfnamefont
  {H.}~\bibnamefont {Zhang}}, \bibinfo {author} {\bibfnamefont {C.~M.}\
  \bibnamefont {Marcus}},\ and\ \bibinfo {author} {\bibfnamefont {C.~J.}\
  \bibnamefont {Palmstrøm}},\ }\bibfield  {title} {\bibinfo {title} {Transport
  studies of epi-al/{InAs} two-dimensional electron gas systems for required
  building-blocks in topological superconductor networks},\ }\href
  {https://doi.org/10.1021/acs.nanolett.9b00494} {\bibfield  {journal}
  {\bibinfo  {journal} {Nano Letters}\ }\textbf {\bibinfo {volume} {19}},\
  \bibinfo {pages} {3083} (\bibinfo {year} {2019})}\BibitemShut {NoStop}%
\bibitem [{\citenamefont {Hell}\ \emph {et~al.}(2017)\citenamefont {Hell},
  \citenamefont {Leijnse},\ and\ \citenamefont
  {Flensberg}}]{hell_two-dimensional_2017}%
  \BibitemOpen
  \bibfield  {author} {\bibinfo {author} {\bibfnamefont {M.}~\bibnamefont
  {Hell}}, \bibinfo {author} {\bibfnamefont {M.}~\bibnamefont {Leijnse}},\ and\
  \bibinfo {author} {\bibfnamefont {K.}~\bibnamefont {Flensberg}},\ }\bibfield
  {title} {\bibinfo {title} {Two-dimensional platform for networks of majorana
  bound states},\ }\href {https://doi.org/10.1103/PhysRevLett.118.107701}
  {\bibfield  {journal} {\bibinfo  {journal} {Physical Review Letters}\
  }\textbf {\bibinfo {volume} {118}},\ \bibinfo {pages} {107701} (\bibinfo
  {year} {2017})}\BibitemShut {NoStop}%
\bibitem [{\citenamefont {Pientka}\ \emph {et~al.}(2017)\citenamefont
  {Pientka}, \citenamefont {Keselman}, \citenamefont {Berg}, \citenamefont
  {Yacoby}, \citenamefont {Stern},\ and\ \citenamefont
  {Halperin}}]{pientka_topological_2017}%
  \BibitemOpen
  \bibfield  {author} {\bibinfo {author} {\bibfnamefont {F.}~\bibnamefont
  {Pientka}}, \bibinfo {author} {\bibfnamefont {A.}~\bibnamefont {Keselman}},
  \bibinfo {author} {\bibfnamefont {E.}~\bibnamefont {Berg}}, \bibinfo {author}
  {\bibfnamefont {A.}~\bibnamefont {Yacoby}}, \bibinfo {author} {\bibfnamefont
  {A.}~\bibnamefont {Stern}},\ and\ \bibinfo {author} {\bibfnamefont {B.~I.}\
  \bibnamefont {Halperin}},\ }\bibfield  {title} {\bibinfo {title} {Topological
  superconductivity in a planar josephson junction},\ }\href
  {https://doi.org/10.1103/PhysRevX.7.021032} {\bibfield  {journal} {\bibinfo
  {journal} {Physical Review X}\ }\textbf {\bibinfo {volume} {7}},\ \bibinfo
  {pages} {021032} (\bibinfo {year} {2017})}\BibitemShut {NoStop}%
\bibitem [{\citenamefont {Dartiailh}\ \emph
  {et~al.}(2021{\natexlab{a}})\citenamefont {Dartiailh}, \citenamefont {Mayer},
  \citenamefont {Yuan}, \citenamefont {Wickramasinghe}, \citenamefont
  {Matos-Abiague}, \citenamefont {Žutić},\ and\ \citenamefont
  {Shabani}}]{dartiailh_phase_2021}%
  \BibitemOpen
  \bibfield  {author} {\bibinfo {author} {\bibfnamefont {M.~C.}\ \bibnamefont
  {Dartiailh}}, \bibinfo {author} {\bibfnamefont {W.}~\bibnamefont {Mayer}},
  \bibinfo {author} {\bibfnamefont {J.}~\bibnamefont {Yuan}}, \bibinfo {author}
  {\bibfnamefont {K.~S.}\ \bibnamefont {Wickramasinghe}}, \bibinfo {author}
  {\bibfnamefont {A.}~\bibnamefont {Matos-Abiague}}, \bibinfo {author}
  {\bibfnamefont {I.}~\bibnamefont {Žutić}},\ and\ \bibinfo {author}
  {\bibfnamefont {J.}~\bibnamefont {Shabani}},\ }\bibfield  {title} {\bibinfo
  {title} {Phase signature of topological transition in josephson junctions},\
  }\href {https://doi.org/10.1103/PhysRevLett.126.036802} {\bibfield  {journal}
  {\bibinfo  {journal} {Physical Review Letters}\ }\textbf {\bibinfo {volume}
  {126}},\ \bibinfo {pages} {036802} (\bibinfo {year}
  {2021}{\natexlab{a}})}\BibitemShut {NoStop}%
\bibitem [{\citenamefont {Ren}\ \emph {et~al.}(2019)\citenamefont {Ren},
  \citenamefont {Pientka}, \citenamefont {Hart}, \citenamefont {Pierce},
  \citenamefont {Kosowsky}, \citenamefont {Lunczer}, \citenamefont {Schlereth},
  \citenamefont {Scharf}, \citenamefont {Hankiewicz}, \citenamefont
  {Molenkamp}, \citenamefont {Halperin},\ and\ \citenamefont
  {Yacoby}}]{ren_topological_2019}%
  \BibitemOpen
  \bibfield  {author} {\bibinfo {author} {\bibfnamefont {H.}~\bibnamefont
  {Ren}}, \bibinfo {author} {\bibfnamefont {F.}~\bibnamefont {Pientka}},
  \bibinfo {author} {\bibfnamefont {S.}~\bibnamefont {Hart}}, \bibinfo {author}
  {\bibfnamefont {A.~T.}\ \bibnamefont {Pierce}}, \bibinfo {author}
  {\bibfnamefont {M.}~\bibnamefont {Kosowsky}}, \bibinfo {author}
  {\bibfnamefont {L.}~\bibnamefont {Lunczer}}, \bibinfo {author} {\bibfnamefont
  {R.}~\bibnamefont {Schlereth}}, \bibinfo {author} {\bibfnamefont
  {B.}~\bibnamefont {Scharf}}, \bibinfo {author} {\bibfnamefont {E.~M.}\
  \bibnamefont {Hankiewicz}}, \bibinfo {author} {\bibfnamefont {L.~W.}\
  \bibnamefont {Molenkamp}}, \bibinfo {author} {\bibfnamefont {B.~I.}\
  \bibnamefont {Halperin}},\ and\ \bibinfo {author} {\bibfnamefont
  {A.}~\bibnamefont {Yacoby}},\ }\bibfield  {title} {\bibinfo {title}
  {Topological superconductivity in a phase-controlled {Josephson} junction},\
  }\href {https://doi.org/10.1038/s41586-019-1148-9} {\bibfield  {journal}
  {\bibinfo  {journal} {Nature}\ }\textbf {\bibinfo {volume} {569}},\ \bibinfo
  {pages} {93} (\bibinfo {year} {2019})}\BibitemShut {NoStop}%
\bibitem [{\citenamefont {Fornieri}\ \emph {et~al.}(2019)\citenamefont
  {Fornieri}, \citenamefont {Whiticar}, \citenamefont {Setiawan}, \citenamefont
  {Portolés}, \citenamefont {Drachmann}, \citenamefont {Keselman},
  \citenamefont {Gronin}, \citenamefont {Thomas}, \citenamefont {Wang},
  \citenamefont {Kallaher}, \citenamefont {Gardner}, \citenamefont {Berg},
  \citenamefont {Manfra}, \citenamefont {Stern}, \citenamefont {Marcus},\ and\
  \citenamefont {Nichele}}]{fornieri_evidence_2019}%
  \BibitemOpen
  \bibfield  {author} {\bibinfo {author} {\bibfnamefont {A.}~\bibnamefont
  {Fornieri}}, \bibinfo {author} {\bibfnamefont {A.~M.}\ \bibnamefont
  {Whiticar}}, \bibinfo {author} {\bibfnamefont {F.}~\bibnamefont {Setiawan}},
  \bibinfo {author} {\bibfnamefont {E.}~\bibnamefont {Portolés}}, \bibinfo
  {author} {\bibfnamefont {A.~C.~C.}\ \bibnamefont {Drachmann}}, \bibinfo
  {author} {\bibfnamefont {A.}~\bibnamefont {Keselman}}, \bibinfo {author}
  {\bibfnamefont {S.}~\bibnamefont {Gronin}}, \bibinfo {author} {\bibfnamefont
  {C.}~\bibnamefont {Thomas}}, \bibinfo {author} {\bibfnamefont
  {T.}~\bibnamefont {Wang}}, \bibinfo {author} {\bibfnamefont {R.}~\bibnamefont
  {Kallaher}}, \bibinfo {author} {\bibfnamefont {G.~C.}\ \bibnamefont
  {Gardner}}, \bibinfo {author} {\bibfnamefont {E.}~\bibnamefont {Berg}},
  \bibinfo {author} {\bibfnamefont {M.~J.}\ \bibnamefont {Manfra}}, \bibinfo
  {author} {\bibfnamefont {A.}~\bibnamefont {Stern}}, \bibinfo {author}
  {\bibfnamefont {C.~M.}\ \bibnamefont {Marcus}},\ and\ \bibinfo {author}
  {\bibfnamefont {F.}~\bibnamefont {Nichele}},\ }\bibfield  {title} {\bibinfo
  {title} {Evidence of topological superconductivity in planar {Josephson}
  junctions},\ }\href {https://doi.org/10.1038/s41586-019-1068-8} {\bibfield
  {journal} {\bibinfo  {journal} {Nature}\ }\textbf {\bibinfo {volume} {569}},\
  \bibinfo {pages} {89} (\bibinfo {year} {2019})}\BibitemShut {NoStop}%
\bibitem [{\citenamefont {Alicea}\ \emph {et~al.}(2011)\citenamefont {Alicea},
  \citenamefont {Oreg}, \citenamefont {Refael}, \citenamefont {von Oppen},\
  and\ \citenamefont {Fisher}}]{alicea_non-abelian_2011}%
  \BibitemOpen
  \bibfield  {author} {\bibinfo {author} {\bibfnamefont {J.}~\bibnamefont
  {Alicea}}, \bibinfo {author} {\bibfnamefont {Y.}~\bibnamefont {Oreg}},
  \bibinfo {author} {\bibfnamefont {G.}~\bibnamefont {Refael}}, \bibinfo
  {author} {\bibfnamefont {F.}~\bibnamefont {von Oppen}},\ and\ \bibinfo
  {author} {\bibfnamefont {M.~P.~A.}\ \bibnamefont {Fisher}},\ }\bibfield
  {title} {\bibinfo {title} {Non-{Abelian} statistics and topological quantum
  information processing in {1D} wire networks},\ }\href
  {https://doi.org/10.1038/nphys1915} {\bibfield  {journal} {\bibinfo
  {journal} {Nature Physics}\ }\textbf {\bibinfo {volume} {7}},\ \bibinfo
  {pages} {412} (\bibinfo {year} {2011})}\BibitemShut {NoStop}%
\bibitem [{\citenamefont {Dartiailh}\ \emph
  {et~al.}(2021{\natexlab{b}})\citenamefont {Dartiailh}, \citenamefont
  {Cuozzo}, \citenamefont {Elfeky}, \citenamefont {Mayer}, \citenamefont
  {Yuan}, \citenamefont {Wickramasinghe}, \citenamefont {Rossi},\ and\
  \citenamefont {Shabani}}]{dartiailh_missing_2021}%
  \BibitemOpen
  \bibfield  {author} {\bibinfo {author} {\bibfnamefont {M.~C.}\ \bibnamefont
  {Dartiailh}}, \bibinfo {author} {\bibfnamefont {J.~J.}\ \bibnamefont
  {Cuozzo}}, \bibinfo {author} {\bibfnamefont {B.~H.}\ \bibnamefont {Elfeky}},
  \bibinfo {author} {\bibfnamefont {W.}~\bibnamefont {Mayer}}, \bibinfo
  {author} {\bibfnamefont {J.}~\bibnamefont {Yuan}}, \bibinfo {author}
  {\bibfnamefont {K.~S.}\ \bibnamefont {Wickramasinghe}}, \bibinfo {author}
  {\bibfnamefont {E.}~\bibnamefont {Rossi}},\ and\ \bibinfo {author}
  {\bibfnamefont {J.}~\bibnamefont {Shabani}},\ }\bibfield  {title} {\bibinfo
  {title} {Missing shapiro steps in topologically trivial josephson junction on
  {InAs} quantum well},\ }\href {https://doi.org/10.1038/s41467-020-20382-y}
  {\bibfield  {journal} {\bibinfo  {journal} {Nature Communications}\ }\textbf
  {\bibinfo {volume} {12}},\ \bibinfo {pages} {78} (\bibinfo {year}
  {2021}{\natexlab{b}})}\BibitemShut {NoStop}%
\bibitem [{\citenamefont {Dynes}\ and\ \citenamefont
  {Fulton}(1971)}]{dynes_supercurrent_1971}%
  \BibitemOpen
  \bibfield  {author} {\bibinfo {author} {\bibfnamefont {R.~C.}\ \bibnamefont
  {Dynes}}\ and\ \bibinfo {author} {\bibfnamefont {T.~A.}\ \bibnamefont
  {Fulton}},\ }\bibfield  {title} {\bibinfo {title} {Supercurrent density
  distribution in josephson junctions},\ }\href
  {https://doi.org/10.1103/PhysRevB.3.3015} {\bibfield  {journal} {\bibinfo
  {journal} {Physical Review B}\ }\textbf {\bibinfo {volume} {3}},\ \bibinfo
  {pages} {3015} (\bibinfo {year} {1971})}\BibitemShut {NoStop}%
\bibitem [{\citenamefont {Wickramasinghe}\ \emph {et~al.}(2018)\citenamefont
  {Wickramasinghe}, \citenamefont {Mayer}, \citenamefont {Yuan}, \citenamefont
  {Nguyen}, \citenamefont {Jiao}, \citenamefont {Manucharyan},\ and\
  \citenamefont {Shabani}}]{wickramasinghe_transport_2018}%
  \BibitemOpen
  \bibfield  {author} {\bibinfo {author} {\bibfnamefont {K.~S.}\ \bibnamefont
  {Wickramasinghe}}, \bibinfo {author} {\bibfnamefont {W.}~\bibnamefont
  {Mayer}}, \bibinfo {author} {\bibfnamefont {J.}~\bibnamefont {Yuan}},
  \bibinfo {author} {\bibfnamefont {T.}~\bibnamefont {Nguyen}}, \bibinfo
  {author} {\bibfnamefont {L.}~\bibnamefont {Jiao}}, \bibinfo {author}
  {\bibfnamefont {V.}~\bibnamefont {Manucharyan}},\ and\ \bibinfo {author}
  {\bibfnamefont {J.}~\bibnamefont {Shabani}},\ }\bibfield  {title} {\bibinfo
  {title} {Transport properties of near surface {InAs} two-dimensional
  heterostructures},\ }\href {https://doi.org/10.1063/1.5050413} {\bibfield
  {journal} {\bibinfo  {journal} {Applied Physics Letters}\ }\textbf {\bibinfo
  {volume} {113}},\ \bibinfo {pages} {262104} (\bibinfo {year}
  {2018})}\BibitemShut {NoStop}%
\bibitem [{\citenamefont {Bauer}\ \emph {et~al.}(2018)\citenamefont {Bauer},
  \citenamefont {Karzig}, \citenamefont {Mishmash}, \citenamefont {Antipov},\
  and\ \citenamefont {Alicea}}]{bauer_dynamics_2018}%
  \BibitemOpen
  \bibfield  {author} {\bibinfo {author} {\bibfnamefont {B.}~\bibnamefont
  {Bauer}}, \bibinfo {author} {\bibfnamefont {T.}~\bibnamefont {Karzig}},
  \bibinfo {author} {\bibfnamefont {R.}~\bibnamefont {Mishmash}}, \bibinfo
  {author} {\bibfnamefont {A.}~\bibnamefont {Antipov}},\ and\ \bibinfo {author}
  {\bibfnamefont {J.}~\bibnamefont {Alicea}},\ }\bibfield  {title} {\bibinfo
  {title} {Dynamics of majorana-based qubits operated with an array of tunable
  gates},\ }\href {https://doi.org/10.21468/SciPostPhys.5.1.004} {\bibfield
  {journal} {\bibinfo  {journal} {{SciPost} Physics}\ }\textbf {\bibinfo
  {volume} {5}},\ \bibinfo {pages} {004} (\bibinfo {year} {2018})}\BibitemShut
  {NoStop}%
\bibitem [{\citenamefont {Zhou}\ \emph {et~al.}(2021)\citenamefont {Zhou},
  \citenamefont {Dartiailh}, \citenamefont {Sardashti}, \citenamefont {Han},
  \citenamefont {Matos-Abiague}, \citenamefont {Shabani},\ and\ \citenamefont
  {Zutic}}]{zhou_fusion_2021}%
  \BibitemOpen
  \bibfield  {author} {\bibinfo {author} {\bibfnamefont {T.}~\bibnamefont
  {Zhou}}, \bibinfo {author} {\bibfnamefont {M.~C.}\ \bibnamefont {Dartiailh}},
  \bibinfo {author} {\bibfnamefont {K.}~\bibnamefont {Sardashti}}, \bibinfo
  {author} {\bibfnamefont {J.~E.}\ \bibnamefont {Han}}, \bibinfo {author}
  {\bibfnamefont {A.}~\bibnamefont {Matos-Abiague}}, \bibinfo {author}
  {\bibfnamefont {J.}~\bibnamefont {Shabani}},\ and\ \bibinfo {author}
  {\bibfnamefont {I.}~\bibnamefont {Zutic}},\ }\bibfield  {title} {\bibinfo
  {title} {Fusion of majorana bound states with mini-gate control in
  two-dimensional systems},\ }\href {http://arxiv.org/abs/2101.09272}
  {\bibfield  {journal} {\bibinfo  {journal} {{arXiv}:2101.09272 [cond-mat]}\ }
  (\bibinfo {year} {2021})},\ \Eprint {https://arxiv.org/abs/2101.09272}
  {2101.09272} \BibitemShut {NoStop}%
\bibitem [{\citenamefont {Suominen}\ \emph {et~al.}(2017)\citenamefont
  {Suominen}, \citenamefont {Danon}, \citenamefont {Kjaergaard}, \citenamefont
  {Flensberg}, \citenamefont {Shabani}, \citenamefont {Palmstrøm},
  \citenamefont {Nichele},\ and\ \citenamefont
  {Marcus}}]{suominen_anomalous_2017}%
  \BibitemOpen
  \bibfield  {author} {\bibinfo {author} {\bibfnamefont {H.~J.}\ \bibnamefont
  {Suominen}}, \bibinfo {author} {\bibfnamefont {J.}~\bibnamefont {Danon}},
  \bibinfo {author} {\bibfnamefont {M.}~\bibnamefont {Kjaergaard}}, \bibinfo
  {author} {\bibfnamefont {K.}~\bibnamefont {Flensberg}}, \bibinfo {author}
  {\bibfnamefont {J.}~\bibnamefont {Shabani}}, \bibinfo {author} {\bibfnamefont
  {C.~J.}\ \bibnamefont {Palmstrøm}}, \bibinfo {author} {\bibfnamefont
  {F.}~\bibnamefont {Nichele}},\ and\ \bibinfo {author} {\bibfnamefont {C.~M.}\
  \bibnamefont {Marcus}},\ }\bibfield  {title} {\bibinfo {title} {Anomalous
  {Fraunhofer} interference in epitaxial superconductor-semiconductor
  {Josephson} junctions},\ }\href {https://doi.org/10.1103/PhysRevB.95.035307}
  {\bibfield  {journal} {\bibinfo  {journal} {Physical Review B}\ }\textbf
  {\bibinfo {volume} {95}},\ \bibinfo {pages} {035307} (\bibinfo {year}
  {2017})}\BibitemShut {NoStop}%
\bibitem [{\citenamefont {Ghatak}\ \emph {et~al.}(2018)\citenamefont {Ghatak},
  \citenamefont {Breunig}, \citenamefont {Yang}, \citenamefont {Wang},
  \citenamefont {Taskin},\ and\ \citenamefont {Ando}}]{ghatak_anomalous_2018}%
  \BibitemOpen
  \bibfield  {author} {\bibinfo {author} {\bibfnamefont {S.}~\bibnamefont
  {Ghatak}}, \bibinfo {author} {\bibfnamefont {O.}~\bibnamefont {Breunig}},
  \bibinfo {author} {\bibfnamefont {F.}~\bibnamefont {Yang}}, \bibinfo {author}
  {\bibfnamefont {Z.}~\bibnamefont {Wang}}, \bibinfo {author} {\bibfnamefont
  {A.~A.}\ \bibnamefont {Taskin}},\ and\ \bibinfo {author} {\bibfnamefont
  {Y.}~\bibnamefont {Ando}},\ }\bibfield  {title} {\bibinfo {title} {Anomalous
  {Fraunhofer} {Patterns} in {Gated} {Josephson} {Junctions} {Based} on the
  {Bulk}-{Insulating} {Topological} {Insulator} {BiSbTeSe} $_{\textrm{2}}$},\
  }\href {https://doi.org/10.1021/acs.nanolett.8b02029} {\bibfield  {journal}
  {\bibinfo  {journal} {Nano Letters}\ }\textbf {\bibinfo {volume} {18}},\
  \bibinfo {pages} {5124} (\bibinfo {year} {2018})}\BibitemShut {NoStop}%
\bibitem [{\citenamefont {Potter}\ and\ \citenamefont
  {Fu}(2013)}]{potter_anomalous_2013}%
  \BibitemOpen
  \bibfield  {author} {\bibinfo {author} {\bibfnamefont {A.~C.}\ \bibnamefont
  {Potter}}\ and\ \bibinfo {author} {\bibfnamefont {L.}~\bibnamefont {Fu}},\
  }\bibfield  {title} {\bibinfo {title} {Anomalous supercurrent from {Majorana}
  states in topological insulator {Josephson} junctions},\ }\href
  {https://doi.org/10.1103/PhysRevB.88.121109} {\bibfield  {journal} {\bibinfo
  {journal} {Physical Review B}\ }\textbf {\bibinfo {volume} {88}},\ \bibinfo
  {pages} {121109} (\bibinfo {year} {2013})}\BibitemShut {NoStop}%
\bibitem [{\citenamefont {Kurter}\ \emph {et~al.}(2015)\citenamefont {Kurter},
  \citenamefont {Finck}, \citenamefont {Hor},\ and\ \citenamefont
  {Van~Harlingen}}]{kurter_evidence_2015}%
  \BibitemOpen
  \bibfield  {author} {\bibinfo {author} {\bibfnamefont {C.}~\bibnamefont
  {Kurter}}, \bibinfo {author} {\bibfnamefont {A.}~\bibnamefont {Finck}},
  \bibinfo {author} {\bibfnamefont {Y.~S.}\ \bibnamefont {Hor}},\ and\ \bibinfo
  {author} {\bibfnamefont {D.~J.}\ \bibnamefont {Van~Harlingen}},\ }\bibfield
  {title} {\bibinfo {title} {Evidence for an anomalous current–phase relation
  in topological insulator josephson junctions},\ }\href
  {https://doi.org/10.1038/ncomms8130} {\bibfield  {journal} {\bibinfo
  {journal} {Nature Communications}\ }\textbf {\bibinfo {volume} {6}},\
  \bibinfo {pages} {7130} (\bibinfo {year} {2015})}\BibitemShut {NoStop}%
\bibitem [{\citenamefont {Williams}\ \emph {et~al.}(2012)\citenamefont
  {Williams}, \citenamefont {Bestwick}, \citenamefont {Gallagher},
  \citenamefont {Hong}, \citenamefont {Cui}, \citenamefont {Bleich},
  \citenamefont {Analytis}, \citenamefont {Fisher},\ and\ \citenamefont
  {Goldhaber-Gordon}}]{williams_unconventional_2012}%
  \BibitemOpen
  \bibfield  {author} {\bibinfo {author} {\bibfnamefont {J.~R.}\ \bibnamefont
  {Williams}}, \bibinfo {author} {\bibfnamefont {A.~J.}\ \bibnamefont
  {Bestwick}}, \bibinfo {author} {\bibfnamefont {P.}~\bibnamefont {Gallagher}},
  \bibinfo {author} {\bibfnamefont {S.~S.}\ \bibnamefont {Hong}}, \bibinfo
  {author} {\bibfnamefont {Y.}~\bibnamefont {Cui}}, \bibinfo {author}
  {\bibfnamefont {A.~S.}\ \bibnamefont {Bleich}}, \bibinfo {author}
  {\bibfnamefont {J.~G.}\ \bibnamefont {Analytis}}, \bibinfo {author}
  {\bibfnamefont {I.~R.}\ \bibnamefont {Fisher}},\ and\ \bibinfo {author}
  {\bibfnamefont {D.}~\bibnamefont {Goldhaber-Gordon}},\ }\bibfield  {title}
  {\bibinfo {title} {Unconventional {Josephson} {Effect} in {Hybrid}
  {Superconductor}-{Topological} {Insulator} {Devices}},\ }\href
  {https://doi.org/10.1103/PhysRevLett.109.056803} {\bibfield  {journal}
  {\bibinfo  {journal} {Physical Review Letters}\ }\textbf {\bibinfo {volume}
  {109}},\ \bibinfo {pages} {056803} (\bibinfo {year} {2012})}\BibitemShut
  {NoStop}%
\bibitem [{\citenamefont {Hui}\ \emph {et~al.}(2014)\citenamefont {Hui},
  \citenamefont {Lobos}, \citenamefont {Sau},\ and\ \citenamefont
  {Das~Sarma}}]{hui_proximity-induced_2014}%
  \BibitemOpen
  \bibfield  {author} {\bibinfo {author} {\bibfnamefont {H.-Y.}\ \bibnamefont
  {Hui}}, \bibinfo {author} {\bibfnamefont {A.~M.}\ \bibnamefont {Lobos}},
  \bibinfo {author} {\bibfnamefont {J.~D.}\ \bibnamefont {Sau}},\ and\ \bibinfo
  {author} {\bibfnamefont {S.}~\bibnamefont {Das~Sarma}},\ }\bibfield  {title}
  {\bibinfo {title} {Proximity-induced superconductivity and {Josephson}
  critical current in quantum spin {Hall} systems},\ }\href
  {https://doi.org/10.1103/PhysRevB.90.224517} {\bibfield  {journal} {\bibinfo
  {journal} {Physical Review B}\ }\textbf {\bibinfo {volume} {90}},\ \bibinfo
  {pages} {224517} (\bibinfo {year} {2014})}\BibitemShut {NoStop}%
\bibitem [{\citenamefont {Albrecht}\ \emph {et~al.}(2016)\citenamefont
  {Albrecht}, \citenamefont {Higginbotham}, \citenamefont {Madsen},
  \citenamefont {Kuemmeth}, \citenamefont {Jespersen}, \citenamefont {Nygård},
  \citenamefont {Krogstrup},\ and\ \citenamefont
  {Marcus}}]{albrecht_exponential_2016}%
  \BibitemOpen
  \bibfield  {author} {\bibinfo {author} {\bibfnamefont {S.~M.}\ \bibnamefont
  {Albrecht}}, \bibinfo {author} {\bibfnamefont {A.~P.}\ \bibnamefont
  {Higginbotham}}, \bibinfo {author} {\bibfnamefont {M.}~\bibnamefont
  {Madsen}}, \bibinfo {author} {\bibfnamefont {F.}~\bibnamefont {Kuemmeth}},
  \bibinfo {author} {\bibfnamefont {T.~S.}\ \bibnamefont {Jespersen}}, \bibinfo
  {author} {\bibfnamefont {J.}~\bibnamefont {Nygård}}, \bibinfo {author}
  {\bibfnamefont {P.}~\bibnamefont {Krogstrup}},\ and\ \bibinfo {author}
  {\bibfnamefont {C.~M.}\ \bibnamefont {Marcus}},\ }\bibfield  {title}
  {\bibinfo {title} {Exponential protection of zero modes in {Majorana}
  islands},\ }\href {https://doi.org/10.1038/nature17162} {\bibfield  {journal}
  {\bibinfo  {journal} {Nature}\ }\textbf {\bibinfo {volume} {531}},\ \bibinfo
  {pages} {206} (\bibinfo {year} {2016})}\BibitemShut {NoStop}%
\bibitem [{\citenamefont {Hasan}\ and\ \citenamefont
  {Kane}(2010)}]{hasan_colloquium_2010}%
  \BibitemOpen
  \bibfield  {author} {\bibinfo {author} {\bibfnamefont {M.~Z.}\ \bibnamefont
  {Hasan}}\ and\ \bibinfo {author} {\bibfnamefont {C.~L.}\ \bibnamefont
  {Kane}},\ }\bibfield  {title} {\bibinfo {title} {Colloquium: {Topological}
  insulators},\ }\href {https://doi.org/10.1103/RevModPhys.82.3045} {\bibfield
  {journal} {\bibinfo  {journal} {Reviews of Modern Physics}\ }\textbf
  {\bibinfo {volume} {82}},\ \bibinfo {pages} {3045} (\bibinfo {year}
  {2010})}\BibitemShut {NoStop}%
\bibitem [{\citenamefont {Qi}\ and\ \citenamefont
  {Zhang}(2011)}]{qi_topological_2011}%
  \BibitemOpen
  \bibfield  {author} {\bibinfo {author} {\bibfnamefont {X.-L.}\ \bibnamefont
  {Qi}}\ and\ \bibinfo {author} {\bibfnamefont {S.-C.}\ \bibnamefont {Zhang}},\
  }\bibfield  {title} {\bibinfo {title} {Topological insulators and
  superconductors},\ }\href {https://doi.org/10.1103/RevModPhys.83.1057}
  {\bibfield  {journal} {\bibinfo  {journal} {Reviews of Modern Physics}\
  }\textbf {\bibinfo {volume} {83}},\ \bibinfo {pages} {1057} (\bibinfo {year}
  {2011})}\BibitemShut {NoStop}%
\bibitem [{\citenamefont {de~Vries}\ \emph {et~al.}(2018)\citenamefont
  {de~Vries}, \citenamefont {Timmerman}, \citenamefont {Ostroukh},
  \citenamefont {van Veen}, \citenamefont {Beukman}, \citenamefont {Qu},
  \citenamefont {Wimmer}, \citenamefont {Nguyen}, \citenamefont {Kiselev},
  \citenamefont {Yi}, \citenamefont {Sokolich}, \citenamefont {Manfra},
  \citenamefont {Marcus},\ and\ \citenamefont {Kouwenhoven}}]{de_vries_h_2018}%
  \BibitemOpen
  \bibfield  {author} {\bibinfo {author} {\bibfnamefont {F.~K.}\ \bibnamefont
  {de~Vries}}, \bibinfo {author} {\bibfnamefont {T.}~\bibnamefont {Timmerman}},
  \bibinfo {author} {\bibfnamefont {V.~P.}\ \bibnamefont {Ostroukh}}, \bibinfo
  {author} {\bibfnamefont {J.}~\bibnamefont {van Veen}}, \bibinfo {author}
  {\bibfnamefont {A.~J.}\ \bibnamefont {Beukman}}, \bibinfo {author}
  {\bibfnamefont {F.}~\bibnamefont {Qu}}, \bibinfo {author} {\bibfnamefont
  {M.}~\bibnamefont {Wimmer}}, \bibinfo {author} {\bibfnamefont {B.-M.}\
  \bibnamefont {Nguyen}}, \bibinfo {author} {\bibfnamefont {A.~A.}\
  \bibnamefont {Kiselev}}, \bibinfo {author} {\bibfnamefont {W.}~\bibnamefont
  {Yi}}, \bibinfo {author} {\bibfnamefont {M.}~\bibnamefont {Sokolich}},
  \bibinfo {author} {\bibfnamefont {M.~J.}\ \bibnamefont {Manfra}}, \bibinfo
  {author} {\bibfnamefont {C.~M.}\ \bibnamefont {Marcus}},\ and\ \bibinfo
  {author} {\bibfnamefont {L.~P.}\ \bibnamefont {Kouwenhoven}},\ }\bibfield
  {title} {\bibinfo {title} {h / e superconducting quantum interference through
  trivial edge states in {InAs}},\ }\href
  {https://doi.org/10.1103/PhysRevLett.120.047702} {\bibfield  {journal}
  {\bibinfo  {journal} {Physical Review Letters}\ }\textbf {\bibinfo {volume}
  {120}},\ \bibinfo {pages} {047702} (\bibinfo {year} {2018})}\BibitemShut
  {NoStop}%
\bibitem [{\citenamefont {de~Vries}\ \emph {et~al.}(2019)\citenamefont
  {de~Vries}, \citenamefont {Sol}, \citenamefont {Gazibegovic}, \citenamefont
  {Veld}, \citenamefont {Balk}, \citenamefont {Car}, \citenamefont {Bakkers},
  \citenamefont {Kouwenhoven},\ and\ \citenamefont
  {Shen}}]{deVries_insb_flake_2019}%
  \BibitemOpen
  \bibfield  {author} {\bibinfo {author} {\bibfnamefont {F.~K.}\ \bibnamefont
  {de~Vries}}, \bibinfo {author} {\bibfnamefont {M.~L.}\ \bibnamefont {Sol}},
  \bibinfo {author} {\bibfnamefont {S.}~\bibnamefont {Gazibegovic}}, \bibinfo
  {author} {\bibfnamefont {R.~L. M. o.~h.}\ \bibnamefont {Veld}}, \bibinfo
  {author} {\bibfnamefont {S.~C.}\ \bibnamefont {Balk}}, \bibinfo {author}
  {\bibfnamefont {D.}~\bibnamefont {Car}}, \bibinfo {author} {\bibfnamefont
  {E.~P. A.~M.}\ \bibnamefont {Bakkers}}, \bibinfo {author} {\bibfnamefont
  {L.~P.}\ \bibnamefont {Kouwenhoven}},\ and\ \bibinfo {author} {\bibfnamefont
  {J.}~\bibnamefont {Shen}},\ }\bibfield  {title} {\bibinfo {title} {Crossed
  andreev reflection in insb flake josephson junctions},\ }\href
  {https://doi.org/10.1103/PhysRevResearch.1.032031} {\bibfield  {journal}
  {\bibinfo  {journal} {Phys. Rev. Research}\ }\textbf {\bibinfo {volume}
  {1}},\ \bibinfo {pages} {032031} (\bibinfo {year} {2019})}\BibitemShut
  {NoStop}%
\bibitem [{\citenamefont {Hart}\ \emph {et~al.}(2014)\citenamefont {Hart},
  \citenamefont {Ren}, \citenamefont {Wagner}, \citenamefont {Leubner},
  \citenamefont {Mühlbauer}, \citenamefont {Brüne}, \citenamefont {Buhmann},
  \citenamefont {Molenkamp},\ and\ \citenamefont {Yacoby}}]{hart_induced_2014}%
  \BibitemOpen
  \bibfield  {author} {\bibinfo {author} {\bibfnamefont {S.}~\bibnamefont
  {Hart}}, \bibinfo {author} {\bibfnamefont {H.}~\bibnamefont {Ren}}, \bibinfo
  {author} {\bibfnamefont {T.}~\bibnamefont {Wagner}}, \bibinfo {author}
  {\bibfnamefont {P.}~\bibnamefont {Leubner}}, \bibinfo {author} {\bibfnamefont
  {M.}~\bibnamefont {Mühlbauer}}, \bibinfo {author} {\bibfnamefont
  {C.}~\bibnamefont {Brüne}}, \bibinfo {author} {\bibfnamefont
  {H.}~\bibnamefont {Buhmann}}, \bibinfo {author} {\bibfnamefont {L.~W.}\
  \bibnamefont {Molenkamp}},\ and\ \bibinfo {author} {\bibfnamefont
  {A.}~\bibnamefont {Yacoby}},\ }\bibfield  {title} {\bibinfo {title} {Induced
  superconductivity in the quantum spin hall edge},\ }\href
  {https://doi.org/10.1038/nphys3036} {\bibfield  {journal} {\bibinfo
  {journal} {Nature Physics}\ }\textbf {\bibinfo {volume} {10}},\ \bibinfo
  {pages} {638} (\bibinfo {year} {2014})}\BibitemShut {NoStop}%
\bibitem [{\citenamefont {Ying}\ \emph {et~al.}(2020)\citenamefont {Ying},
  \citenamefont {He}, \citenamefont {Yang}, \citenamefont {Liu}, \citenamefont
  {Lyu}, \citenamefont {Zhang}, \citenamefont {Liu}, \citenamefont {Zhao},
  \citenamefont {Jiang}, \citenamefont {Ji}, \citenamefont {Fan}, \citenamefont
  {Yang}, \citenamefont {Jing}, \citenamefont {Liu}, \citenamefont {Cao},
  \citenamefont {Wang}, \citenamefont {Lu},\ and\ \citenamefont
  {Qu}}]{ying_magnitude_2020}%
  \BibitemOpen
  \bibfield  {author} {\bibinfo {author} {\bibfnamefont {J.}~\bibnamefont
  {Ying}}, \bibinfo {author} {\bibfnamefont {J.}~\bibnamefont {He}}, \bibinfo
  {author} {\bibfnamefont {G.}~\bibnamefont {Yang}}, \bibinfo {author}
  {\bibfnamefont {M.}~\bibnamefont {Liu}}, \bibinfo {author} {\bibfnamefont
  {Z.}~\bibnamefont {Lyu}}, \bibinfo {author} {\bibfnamefont {X.}~\bibnamefont
  {Zhang}}, \bibinfo {author} {\bibfnamefont {H.}~\bibnamefont {Liu}}, \bibinfo
  {author} {\bibfnamefont {K.}~\bibnamefont {Zhao}}, \bibinfo {author}
  {\bibfnamefont {R.}~\bibnamefont {Jiang}}, \bibinfo {author} {\bibfnamefont
  {Z.}~\bibnamefont {Ji}}, \bibinfo {author} {\bibfnamefont {J.}~\bibnamefont
  {Fan}}, \bibinfo {author} {\bibfnamefont {C.}~\bibnamefont {Yang}}, \bibinfo
  {author} {\bibfnamefont {X.}~\bibnamefont {Jing}}, \bibinfo {author}
  {\bibfnamefont {G.}~\bibnamefont {Liu}}, \bibinfo {author} {\bibfnamefont
  {X.}~\bibnamefont {Cao}}, \bibinfo {author} {\bibfnamefont {X.}~\bibnamefont
  {Wang}}, \bibinfo {author} {\bibfnamefont {L.}~\bibnamefont {Lu}},\ and\
  \bibinfo {author} {\bibfnamefont {F.}~\bibnamefont {Qu}},\ }\bibfield
  {title} {\bibinfo {title} {Magnitude and spatial distribution control of the
  supercurrent in bi $_{\textrm{2}}$ o $_{\textrm{2}}$ se-based josephson
  junction},\ }\href {https://doi.org/10.1021/acs.nanolett.0c00025} {\bibfield
  {journal} {\bibinfo  {journal} {Nano Letters}\ }\textbf {\bibinfo {volume}
  {20}},\ \bibinfo {pages} {2569} (\bibinfo {year} {2020})}\BibitemShut
  {NoStop}%
\bibitem [{\citenamefont {Huang}\ \emph {et~al.}(2020)\citenamefont {Huang},
  \citenamefont {Narayan}, \citenamefont {Zhang}, \citenamefont {Xie},
  \citenamefont {Ai}, \citenamefont {Liu}, \citenamefont {Yi}, \citenamefont
  {Shi}, \citenamefont {Sanvito},\ and\ \citenamefont {Xiu}}]{huang_edge_2020}%
  \BibitemOpen
  \bibfield  {author} {\bibinfo {author} {\bibfnamefont {C.}~\bibnamefont
  {Huang}}, \bibinfo {author} {\bibfnamefont {A.}~\bibnamefont {Narayan}},
  \bibinfo {author} {\bibfnamefont {E.}~\bibnamefont {Zhang}}, \bibinfo
  {author} {\bibfnamefont {X.}~\bibnamefont {Xie}}, \bibinfo {author}
  {\bibfnamefont {L.}~\bibnamefont {Ai}}, \bibinfo {author} {\bibfnamefont
  {S.}~\bibnamefont {Liu}}, \bibinfo {author} {\bibfnamefont {C.}~\bibnamefont
  {Yi}}, \bibinfo {author} {\bibfnamefont {Y.}~\bibnamefont {Shi}}, \bibinfo
  {author} {\bibfnamefont {S.}~\bibnamefont {Sanvito}},\ and\ \bibinfo {author}
  {\bibfnamefont {F.}~\bibnamefont {Xiu}},\ }\bibfield  {title} {\bibinfo
  {title} {Edge superconductivity in multilayer {WTe}2 josephson junction},\
  }\href {https://doi.org/10.1093/nsr/nwaa114} {\bibfield  {journal} {\bibinfo
  {journal} {National Science Review}\ }\textbf {\bibinfo {volume} {7}},\
  \bibinfo {pages} {1468} (\bibinfo {year} {2020})}\BibitemShut {NoStop}%
\bibitem [{\citenamefont {Zhu}\ \emph {et~al.}(2017)\citenamefont {Zhu},
  \citenamefont {Kretinin}, \citenamefont {Thompson}, \citenamefont {Bandurin},
  \citenamefont {Hu}, \citenamefont {Yu}, \citenamefont {Birkbeck},
  \citenamefont {Mishchenko}, \citenamefont {Vera-Marun}, \citenamefont
  {Watanabe}, \citenamefont {Taniguchi}, \citenamefont {Polini}, \citenamefont
  {Prance}, \citenamefont {Novoselov}, \citenamefont {Geim},\ and\
  \citenamefont {Ben~Shalom}}]{zhu_edge_2017}%
  \BibitemOpen
  \bibfield  {author} {\bibinfo {author} {\bibfnamefont {M.~J.}\ \bibnamefont
  {Zhu}}, \bibinfo {author} {\bibfnamefont {A.~V.}\ \bibnamefont {Kretinin}},
  \bibinfo {author} {\bibfnamefont {M.~D.}\ \bibnamefont {Thompson}}, \bibinfo
  {author} {\bibfnamefont {D.~A.}\ \bibnamefont {Bandurin}}, \bibinfo {author}
  {\bibfnamefont {S.}~\bibnamefont {Hu}}, \bibinfo {author} {\bibfnamefont
  {G.~L.}\ \bibnamefont {Yu}}, \bibinfo {author} {\bibfnamefont
  {J.}~\bibnamefont {Birkbeck}}, \bibinfo {author} {\bibfnamefont
  {A.}~\bibnamefont {Mishchenko}}, \bibinfo {author} {\bibfnamefont {I.~J.}\
  \bibnamefont {Vera-Marun}}, \bibinfo {author} {\bibfnamefont
  {K.}~\bibnamefont {Watanabe}}, \bibinfo {author} {\bibfnamefont
  {T.}~\bibnamefont {Taniguchi}}, \bibinfo {author} {\bibfnamefont
  {M.}~\bibnamefont {Polini}}, \bibinfo {author} {\bibfnamefont {J.~R.}\
  \bibnamefont {Prance}}, \bibinfo {author} {\bibfnamefont {K.~S.}\
  \bibnamefont {Novoselov}}, \bibinfo {author} {\bibfnamefont {A.~K.}\
  \bibnamefont {Geim}},\ and\ \bibinfo {author} {\bibfnamefont
  {M.}~\bibnamefont {Ben~Shalom}},\ }\bibfield  {title} {\bibinfo {title} {Edge
  currents shunt the insulating bulk in gapped graphene},\ }\href
  {https://doi.org/10.1038/ncomms14552} {\bibfield  {journal} {\bibinfo
  {journal} {Nature Communications}\ }\textbf {\bibinfo {volume} {8}},\
  \bibinfo {pages} {14552} (\bibinfo {year} {2017})}\BibitemShut {NoStop}%
\bibitem [{\citenamefont {Canali}\ \emph {et~al.}(1998)\citenamefont {Canali},
  \citenamefont {Wildöer}, \citenamefont {Kerkhof},\ and\ \citenamefont
  {Kouwenhoven}}]{canali_low-temperature_1998}%
  \BibitemOpen
  \bibfield  {author} {\bibinfo {author} {\bibfnamefont {L.}~\bibnamefont
  {Canali}}, \bibinfo {author} {\bibfnamefont {J.}~\bibnamefont {Wildöer}},
  \bibinfo {author} {\bibfnamefont {O.}~\bibnamefont {Kerkhof}},\ and\ \bibinfo
  {author} {\bibfnamefont {L.}~\bibnamefont {Kouwenhoven}},\ }\bibfield
  {title} {\bibinfo {title} {Low-temperature {STM} on {InAs}(110) accumulation
  surfaces},\ }\href {https://doi.org/10.1007/s003390051111} {\bibfield
  {journal} {\bibinfo  {journal} {Applied Physics A: Materials Science \&
  Processing}\ }\textbf {\bibinfo {volume} {66}},\ \bibinfo {pages} {S113}
  (\bibinfo {year} {1998})}\BibitemShut {NoStop}%
\bibitem [{\citenamefont {Mead}\ and\ \citenamefont
  {Spitzer}(1964)}]{mead_fermi_1964}%
  \BibitemOpen
  \bibfield  {author} {\bibinfo {author} {\bibfnamefont {C.~A.}\ \bibnamefont
  {Mead}}\ and\ \bibinfo {author} {\bibfnamefont {W.~G.}\ \bibnamefont
  {Spitzer}},\ }\bibfield  {title} {\bibinfo {title} {Fermi {Level} {Position}
  at {Metal}-{Semiconductor} {Interfaces}},\ }\href
  {https://doi.org/10.1103/PhysRev.134.A713} {\bibfield  {journal} {\bibinfo
  {journal} {Physical Review}\ }\textbf {\bibinfo {volume} {134}},\ \bibinfo
  {pages} {A713} (\bibinfo {year} {1964})}\BibitemShut {NoStop}%
\bibitem [{\citenamefont {Bhargava}\ \emph {et~al.}(1997)\citenamefont
  {Bhargava}, \citenamefont {Blank}, \citenamefont {Narayanamurti},\ and\
  \citenamefont {Kroemer}}]{bhargava_fermi-level_1997}%
  \BibitemOpen
  \bibfield  {author} {\bibinfo {author} {\bibfnamefont {S.}~\bibnamefont
  {Bhargava}}, \bibinfo {author} {\bibfnamefont {H.-R.}\ \bibnamefont {Blank}},
  \bibinfo {author} {\bibfnamefont {V.}~\bibnamefont {Narayanamurti}},\ and\
  \bibinfo {author} {\bibfnamefont {H.}~\bibnamefont {Kroemer}},\ }\bibfield
  {title} {\bibinfo {title} {Fermi-level pinning position at the {Au}–{InAs}
  interface determined using ballistic electron emission microscopy},\ }\href
  {https://doi.org/10.1063/1.118271} {\bibfield  {journal} {\bibinfo  {journal}
  {Applied Physics Letters}\ }\textbf {\bibinfo {volume} {70}},\ \bibinfo
  {pages} {759} (\bibinfo {year} {1997})}\BibitemShut {NoStop}%
\end{thebibliography}%


\providecommand{\noopsort}[1]{}\providecommand{\singleletter}[1]{#1}%
%

\end{document}